\documentstyle{article}
\begin{document}
\title{STATIONARY STATES IN SATURATED TWO-PHOTON PROCESSES AND
GENERATION OF PHASE-AVERAGED MIXTURES OF EVEN AND ODD QUANTUM STATES}

\author{{V.V. Dodonov}
and
{S.S. Mizrahi}\\
Departamento de F\'{\i}sica, Universidade Federal de S\~ao Carlos,\\
Via Washington Luiz km 235, 13565-905  S\~ao Carlos,  SP,  Brazil}

\date{}
\maketitle

\abstract{
We consider a relaxation of a single mode of the quantized field
in a presence of one- and
two-photon absorption and emission processes.
Exact stationary solutions of the master equation
for the diagonal elements of the density matrix in the Fock basis are found
in the case of completely saturated two-photon emission.
If two-photon processes dominate over single-photon ones,
the stationary state is a mixture of phase averaged even and odd coherent
states.
}

\section{Introduction}

In many cases, the quantum relaxation can be described in
the framework of the master equation for the statistical operator
$\widehat\rho$  [1--3] ($\hbar=1$)
\begin{equation}
 \frac{\partial\widehat \rho }{\partial t}
+i\left[\widehat {H},\,\widehat \rho\right]
= \frac12 \sum_k \left(2\widehat A_k\widehat\rho
\widehat A^{\dagger}_k
-\widehat A^{\dagger}_k\widehat A_k\widehat\rho
-\widehat\rho\widehat A^{\dagger}_k\widehat A_k\right),
\label{master}\end{equation}
the $\widehat A_k$'s ($k=1,2,\cdots$) being some linear
operators. If the system under study is an electromagnetic
field mode (or an equivalent harmonic oscillator), then  Hamiltonian
$\widehat {H}$ and each operator
$\widehat A_k$ can be expressed in terms of the annihilation and
creation operators $\widehat a$, $\widehat a^{\dagger}$ satisfying
the commutation relation
$\left[\widehat a, \widehat a^{\dagger}\right]=1$.
There exists a specific subfamily of master equations, defined by
operators $\widehat A_k$ in the form
\begin{equation}
\widehat A_k =\left[f_k^{(a)}(\widehat n)\right]^{1/2}\widehat a^k
\quad \mbox{or} \quad
\widehat A_k =(\widehat a^{\dagger})^k\left[f_k^{(e)}
(\widehat n)\right]^{1/2},
\label{subfam}
\end{equation}
 $f_k^{(a,e)}(\widehat n)$ being arbitrary nonnegative functions of
the photon number operator $\widehat n=\widehat a^{\dagger}\widehat a$.
If the Hamiltonian is diagonal in the Fock basis,
$\widehat{H}=\widehat{H}(\widehat n)$,
then Eqs (\ref{master}) and (\ref{subfam}) result in
a {\it closed\/} set of equations for the occupation probabilities
$p_n=\langle n\vert\widehat{\rho}\vert n\rangle$ ($n=0,1,\ldots$)
\begin{eqnarray}
&&\dot{p}_n=\sum_k
\left[ n_k^{(+)}f_k^{(a)}(n)p_{n+k} -
n_k^{(-)}f_k^{(a)}(n-k)p_n\right]
\nonumber\\
&&-\sum_k \left[ n_k^{(+)}f_k^{(e)}(n)p_{n} -
n_k^{(-)}f_k^{(e)}(n-k)p_{n-k}\right],
\label{12phot}\end{eqnarray}
where
$ n_k^{(+)}\equiv (n+k)!/n!$,
$n_k^{(-)}\equiv n!/(n-k)!$.
Note that Eq. (\ref{12phot}) does not contain off-diagonal matrix elements.
And vice versa, the evolution of the off-diagonal elements is completely
independent of the evolution of the diagonal ones, since the derivative
$\partial\langle m|\widehat\rho|n\rangle/\partial t$ is expressed in terms
of the elements $\langle m+k|\widehat\rho|n+k\rangle$ only (with
$k=0,\pm 1,\pm 2,\ldots$). This means that the stationary
solutions to Eqs (\ref{master})-(\ref{subfam}) describe completely decoherent
states, since all off-diagonal elements relaxate to zero values.

If $f_k^{(a)}$ and $f_k^{(e)}$ are constant positive numbers,
then the terms labeled with
superscripts $^{(a)}$ or $^{(e)}$ describe the processes of $k$-photon
absorption or emission by some atomic reservoirs \cite{Shen}.
The choice
\begin{equation}
f_k^{(e)}(n)=d_k\left[1+\gamma_k n_k^{(+)}\right]^{-1}, \quad
f_k^{(a)}(n)=const
\label{scul-k}
\end{equation}
corresponds to a multiphoton generalization of the Scully-Lamb
\cite{SL} single-mode laser equation, the coefficient $\gamma_k$ being
responsible for the saturation effect.
Another equation, implying the presence of emission processes of
all orders, described by means of powers of the shift operator
$ \widehat u p_n\equiv p_{n-1}-p_n$, was proposed by Golubev and Sokolov
\cite{GolSok}:
\begin{equation}
\dot{p}_n= r\ln\left(1+\widehat u\right)p_n +
D_1^{(a)}\left[ (n+1)p_{n+1} - n p_n\right].
\label{GS-eq}
\end{equation}
In this case, coefficients $f_k^{(e)}(n)$ are some rational
functions determined by the Taylor expansion of function $\ln(1+ u)$.
A more general equation, with transcendental (trigonometrical)
coefficients $f_k^{(e)}(n)$, was obtained in \cite{Berg}.

It appears that the family of known exact solutions of
Eqs. (\ref{master}) and (\ref{12phot}) is not very large.
For instance, exact solutions
for {\it arbitrary\/} functions $f_k^{(a)}(n)$ and $f_k^{(e)}(n)$
were found only in the {\it stationary\/} case with $k=1$ \cite{Land}.
Exact {\it time dependent\/} solutions of Eq. (\ref{12phot})
(as well as of equations for the off-diagonal elements)
in the case of constant coefficients $f_1^{(a)}$ and $f_1^{(e)}$
were obtained in \cite{Zeld}
(see also \cite{Bar}). Exact time evolution for the one-photon
Scully-Lamb equation without absorption ($f_1^{(a)}=0$)
was found in \cite{Band-Vo}.

For $k\ge 2$ (multiphoton processes), exact {\it time dependent\/} solutions
of Eq. (\ref{12phot}) with a
{\it single\/} nonzero coefficient
(either $f_k^{(a)}$ or $f_k^{(e)}$) were
found in Refs. [11--21].
In particular, the two-photon emission with $f_2^{(e)}=const$
was considered in \cite{Lambrop,McNeil}.
The case of the function $f_2^{(e)}(n)$ in the
modified Scully-Lamb form (\ref{scul-k}) was treated in \cite{Band-Vo}.
The two-photon absorption without emission ($f_2^{(a)}=const$) was
studied in detail in [13--17]. The case
$f_k^{(a)}=const$ with an arbitrary $k\ge2$ was investigated in
\cite{Voigt,Zub}, and the case $f_k^{(e)}=const$ --- in \cite{Zub}
(the evolution of the off-diagonal
matrix elements in the case of two-photon absorption was studied in
\cite{Simaan2}, and for $k$-photon absorption --- in \cite{Zub,Vo-Band}).
Other references can be found, e.g., in \cite{Paul,Perina}.
Exact time dependent solutions with {\it two\/} nonzero coefficients
were obtained in [6,24--26].
In particular, the time dependent absorption problem with constant
coefficients $f_1^{(a)}$ and $f_2^{(a)}$ was
solved in Ref. \cite{Band77} (a more detailed analysis was given
recently in \cite{DM-new}).

A simplified version of equation (\ref{GS-eq}), with the operator
$\ln\left(1+\widehat u\right)$ replaced by the first two terms of the
Taylor expansion, $\widehat u -\widehat u^2/2$, was solved in
\cite{GolSok}, whereas the solution in a general case was given
in \cite{Tan}.
Other exact solutions with two (or more) nonzero coefficients were found
in the stationary regime only.
For the case of simultaneous $k$-photon absorption and $k$-photon emission
(the so called systems in detailed balance) this was done in \cite{McNeil-1}
for the coefficients in the form (\ref{scul-k}),
and in \cite{DM95} for constant $f_k^{(a)}$ and $f_k^{(e)}$
(see also \cite{Simaan} for $k=2$).
A scheme of obtaining exact stationary solutions of the
two-photon Scully-Lamb equation with single-photon losses
($f_1^{(a)}=const$, $f_2^{(a)}=0$) was given in \cite{Zub2}.
It was generalized to an arbitrary $k\ge 2$ in \cite{Band-Vo85}.
A stationary solution the case
$f_2^{(e)}=An/(n+2)$, $f_2^{(a)}=Bn(n-1)$, was found in \cite{Bul} .
The case of {\it three\/} coefficients, $f_{1,2}^{(a)}=const$,
$f_1^{(e)}(n)=const$  or $f_1^{(e)}(n)= A(n+1)^{-1}$, was considered
briefly in \cite{Ban-Ri,Hild}.
A detailed analysis of the problem with three constant
coefficients, $f_1^{(a)}$, $f_2^{(a)}$ and $f_1^{(e)}$
(when one has one- and two-photon absorption,
but only one-photon emission), was given recently in \cite{DMnew2}.

The aim of the present article is to find a
 stationary exact solution to Eq. (\ref{12phot}) in the presence of a
{\it two-photon emisssion\/}. Although we did not succeed in solving the
equation for a {\it constant\/} emission coefficient $f_2^{(e)}$,
we found that the problem
can be solved in the {\it complete saturation regime\/}
($\gamma_2 \gg 1$) of the two-photon Scully-Lamb equation (\ref{scul-k}),
when the two-photon emission
is decribed by the function $f_2^{(e)}(n)=D[(n+1)(n+2)]^{-1}$
(with the standard form $f_k^{(a)}=const$ for the absorption terms, $k=1,2$).
Under this restriction, there exists a 4-parameter family of equations, whose
solutions can be expressed in terms of the confluent hypergeometric function
or its special cases.

The physical motivation for studying the new model
(which is, in turn, a special case of a more general 6-parameter family of
equations admitting exact solutions) is explained by the fact
that in the case of weak one-photon processes the stationary solutions
describe an interesting class of nonclassical states, namely
{\it phase-averaged even and odd states\/} (PAEOS),
which are mixed analogs of the even and odd coherent (pure) states (EOCS)
\begin{equation}
|\alpha_\pm\rangle = N_\pm ( | \alpha\rangle \pm |- \alpha\rangle ),
\quad
N_+^2 = \frac { \exp(| \alpha|^2) } { 4\cosh(| \alpha|^2) }, \quad
N_-^2 = \frac { \exp(| \alpha|^2) } { 4\sinh(| \alpha|^2) }
\label{evod}
\end{equation}
($| \alpha\rangle$ means the Glauber coherent state \cite{Gla}),
 introduced in \cite{DM74} and studied, e.g. in [23,37--41]
(for generalizations see, e.g. [42--46]).
Since EOCS are the simplest examples of the ``Schr\"odinger cat states''
(another simple example is the Yurke-Stoler state \cite{YS}
$ |\widetilde{\alpha}\rangle_{YS}=
\left(| \alpha\rangle +i | -\alpha\rangle\right)/\sqrt2 $, the principal
difference between EOCS and YS-states is that the EOCS have super-
(even states) or sub-Poisson (odd states) photon statistics, whereas the
statistics of the YS-states is exactly Poissonian),
many authors considered different schemes of generating these states in
physical processes: see, e.g. [48--52]
and an extensive review \cite{Buz}.
It is known, in particular \cite{Gerry93}, that even and odd coherent
states can arise due to the competition between a two-photon absorption and
two-photon parametric processes (described by means of a {\it nondiagonal\/}
Hamiltonian $\widehat{H}(t)$) for a special initial field state.
Here we show that one can obtain {\it phase-averaged\/} even and odd states
using a diagonal Hamiltonian, provided that the (saturated) two-photon
emission is admitted.

\section{A family of exact solutions}\label{sec-exact}

A complete information about the distribution $\left\{p_n\right\}$
is contained in the generating function (GF)
$F(z,t)=\sum_{n=0}^{\infty} p_n(t) z^n$.
Its derivatives yield the probabilities $p_n$ and the factorial moments
$ {\cal N}_m \equiv \sum_{n=m}^{\infty}n(n-1)\cdots(n-m+1) p_n $:
\begin{equation}
p_n = \frac1{n!}\left.\frac{\partial^n F}{\partial z^n}\right|_{z=0},
\quad
{\cal N}_m= \left.\frac{\partial^m F}{ \partial z^m} \right|_{z=1}.
\label{p-fac}\end{equation}
If the products $n_k^{(\pm)}f_k^{(a,e)}(n)$ are polynomials of $n$, then
one can replace the infinite system of difference equations (\ref{12phot})
for the probabilities $p_{n}$
by a single differential equation for $F(z)$.
In the simplest cases, corresponding
either to one-photon processes \cite{Zeld,Bar}, or to a specific form of the
emission operator (\ref{GS-eq}) \cite{GolSok,Tan}, one gets
a linear differential equation of the first order.
In the most of other known
cases, the generating functions satisfy the second order
differential equations of the hypergeometric type [13--18,24,26]
(an exception is the case considered in \cite{Bul}, where a specific
equation of the {\it fourth\/} order was solved with the aid of the
Laplace method).
One can verify that the set of {\it stationary\/} ($\dot{p}_n=0$)
equations (\ref{12phot}) results in the {\it second order\/} equation with
{\it linear\/} coefficients ($F^{\prime}\equiv dF/dz$),
\begin{eqnarray}
&&\left[D_{2}^{(a)}(1+ z) + \left( D_{10}^{(a)} + D_{12}^{(a)}
\right) z\right] F^{\prime\prime} \nonumber\\
&&+\left[D_{1}^{(a)} +2D_{12}^{(a)}  - z
\left( D_{1}^{(e)} +\sum_{j\neq 1} W_{1j}^{(e)}\right) \right]F^{\prime}
\nonumber\\
&&-\left[D_{2}^{(e)} (1+ z) +
D_{1}^{(e)} + D_{11}^{(e)} +\sum_{j\neq 1}j W_{1j}^{(e)}
 \right]F=0,
\label{stat-eqgen}\end{eqnarray}
provided that functions $f_k^{(a,e)}$ have the following form:
\begin{equation}
 f_2^{(e)}(n)=\frac{D_{2}^{(e)}}{(n+1)(n+2)}, \qquad
 f_2^{(a)}(n)=D_{2}^{(a)},
\label{f2-ea}
\end{equation}
\begin{equation}
 f_1^{(a)}(n)= D_{1}^{(a)}  + D_{10}^{(a)} n + D_{12}^{(a)}( n+2) ,
\label{f1-a}
\end{equation}
\begin{equation}
 f_1^{(e)}(n)= D_{1}^{(e)} + \frac1{ n+1}\left( D_{11}^{(e)} +
 \sum_{j\neq 1} W_{1j}^{(e)} (n+j)\right)  ,
\label{f1-e}
\end{equation}
$D_{ij}^{(a,e)}$ and $W_{1j}^{(e)}$ being nonnegative constant coefficients,
whereas
$j$ can be any integer (excepting the value $j=1$, which is distinguished
for the sake of convenience, because it corresponds to the usual one-photon
emission described by the coefficient $D_{1}^{(e)}$).
Since Eq. (\ref{stat-eqgen}) can be reduced
to the Kummer equation \cite{Bate}
\begin{equation}
xy^{\prime\prime} +(c-x)y^{\prime} -ay=0,
\label{Kummer}\end{equation}
we have a whole
family of master equations admitting exact stationary solutions in terms
of the confluent hypergeometric function
\begin{equation}
\Phi(a;c;x)=\sum_{k=0}^{\infty}\frac{(a)_k x^k}{(c)_k k!},
\label{def-kum}
\end{equation}
where $ (a)_n\equiv a(a+1)\cdots(a+n-1)$.
This family is determined by 6
independent positive parameters, so it is larger than any one considered
until now. Note, however, that we have some freedom only in the choice of
terms responsible for the one-photon processes, while the structure of
two-photon terms is fixed: the usual two-photon absorption and the completely
saturated two-photon emission (corresponding to the limit
$\gamma_2\to\infty$, $d_2/\gamma_2\to D_{2}^{(e)}$ in Eq. (\ref{scul-k})).

Here we confine ourselves to the special case
$D_{10}^{(a)} = D_{12}^{(a)}=W_{1j}^{(e)}=0$. Then we have 5 independent
parameters, $D_{1,2}^{(a,e)}$ and $D_{11}^{(e)}$. Normalizing all the
coefficients by the two-photon absorption coefficient, $D_{2}^{(a)}$,
we arrive at the following set of stationary equations for the probabilities
and for the generating function:
\begin{eqnarray}
&& \nu \left\{ (n+1)p_{n+1}-np_n
-s\left[(n+1)p_{n}-np_{n-1}\right]
-\sigma\left[p_{n}-p_{n-1}\right]\right\} \nonumber\\
&&+ (n+1)(n+2)p_{n+2}-n(n-1)p_n
-r^2\left(p_{n}-p_{n-2}\right) =0,
\label{p-12eq}\end{eqnarray}
\begin{equation}
(1+ z)  F^{\prime\prime}
+\nu (1 - sz) F^{\prime}
-\left[\nu(s+\sigma) + r^2 (1+ z) \right]F=0,
\label{F-eq12}\end{equation}
where we have introduced the dimensionless coefficients
\begin{equation}
\nu\equiv D_1^{(a)}/D_2^{(a)},\quad s\equiv D_1^{(e)}/D_1^{(a)},
\quad \sigma\equiv D_{11}^{(e)}/D_1^{(a)},
\quad r^2\equiv D_2^{(e)}/D_2^{(a)}.
\label{def-coef}
\end{equation}
A regular solution to Eq. (\ref{F-eq12}) (without a singularity at $z=-1$)
satisfying the normalization condition $F(1)=1$ can be expressed
in terms of the confluent hypergeometric function
\begin{equation}
F(z)=e^{h(1-z)}\frac{\Phi(\nu g\,;\,\nu[1+s]\,;\,R[1+z])}
{\Phi(\nu g\,;\,\nu[1+s]\,;\,2R)},
\label{sol-kum}\end{equation}
where
\[ R=\left[(\nu s)^2 +4r^2\right]^{1/2}, \quad
h=\frac12(R- \nu s), \quad
g=\frac1R[s+\sigma +h(1+s)]. \]
In particular, if $s=0$, then $R=2r$, $h=r$, and $g=\frac12(1+\sigma/r)$.
The probabilities and factorial moments can be found with the aid
of Eq. (\ref{p-fac}) and the relation \cite{Bate}
\[\frac{d^n}{dx^n}\Phi(a;c;x)=\frac{(a)_n}{(c)_n}\Phi(a+n;c+n;x). \]

\subsection{Some special cases}\label{limit}

If the one-photon processes dominate over two-photon ones,
$\nu\to\infty$ (whereas $r,s,\sigma$ remain finite), then $h\to 0$,
$R\approx \nu s$, $\nu g\to 1+\sigma/s$. Replacing $(c)_k$ by $c^k$
in the Kummer series (\ref{def-kum}) for $c\gg 1$, we obtain the following
limit of the generating function (\ref{sol-kum}) as $\nu\to\infty$:
\begin{equation}
F(z)=\left(\frac{1-s}{1-sz}\right)^{1+\sigma/s}.
\label{F-negbin}\end{equation}
This is the GF of the {\it negative binomial distribution},
which was considered in connection with the problems of quantum optics,
e.g. in Refs. \cite{Joshi,Agar92}.
For $\sigma=0$ (\ref{F-negbin}) becomes the GF of the
thermal (Planck's) distribution, whereas for $s\to 0$ it goes to the
GF of the Poisson distribution. Evidently, Eq. (\ref{F-negbin}) is valid
only for $s<1$, whereas the general formula (\ref{sol-kum}) holds for
all nonnegative parameters $s,\sigma,\nu, r$. If $s\ge 1$, the
asymptotic behaviour of GF at $\nu\gg 1$ is more complicated.
For instance, in the particular case $r=\sigma=0$,
the distribution $\{p_n\}$ becomes Gaussian when
$\nu\gg 1$ and $s\ge 1$ \cite{DMnew2}.

The Poisson distribution arises also in the limit $s\to\infty$.
Then the GF tends to $\exp(z-1)$, i.e. the limit
distribution has the mean photon number $\overline{n}=1$, independently
on the values of other (finite) parameters, $\nu,\sigma,r$.

Another simple expression for the GF is obtained in absence of the
two-photon absorption, $D_2^{(a)}=0$. Then instead of Eq. (\ref{F-eq12})
we get the first order equation
\begin{equation}
\nu (1 - sz) F^{\prime}
-\left[\nu(s+\sigma) + \rho (1+ z) \right]F=0
\label{F-eqD2a=0}\end{equation}
($\rho\equiv D_2^{(e)}/D_1^{(a)}$), whose normalized solution reads
\begin{equation}
F(z)=\left(\frac{1-s}{1-sz}\right)^{1+\gamma}
\exp\left[\frac{\rho}{s}(1-z)\right], \quad
\gamma=\frac1s\left(\sigma +\rho +\frac{\rho}{s}\right).
\label{GF-D2a=0}
\end{equation}

\section{Phase-averaged even and odd states}

Now let us consider the situation, when the two-photon processes
dominate over the one-photon counterparts.
Suppose first that the one-photon processes are completely absent, i.e.
$\nu=\nu s=\nu\sigma= 0$. Then the solution of Eq. (\ref{F-eq12}) satisfying
the condition $F(1)=1$ reads
\begin{equation}
F(z)=(1-\beta)\frac{\cosh(rz)}{\cosh(r)} +\beta\frac{\sinh(rz)}{\sinh(r)},
\label{F-evod}
\end{equation}
so the occupation probabilities are given by
\begin{equation}
p_{2k}=\frac{(1-\beta)r^{2k}}{(2k)!\cosh(r)} , \quad
p_{2k+1}=\frac{\beta r^{2k+1}}{(2k+1)!\sinh(r)}.
\label{p-evod}
\end{equation}
The distribution (\ref{p-evod}) is nothing but a combination of the
photon distribution functions of the
{\it even and odd coherent states\/} $|\alpha_+\rangle$ and
$|\alpha_-\rangle$ (\ref{evod})
with relative weights $1-\beta$ and $\beta$, respectively, provided that
$|\alpha|^2$ is identified with the ratio of the two-photon emission and
absorption coefficients $r$.
The relative weight of the odd states $\beta$ is determined
by the initial conditions, since there is no correlation between even
and odd states: $\beta=\sum_{k=0}^{\infty} p_{2k+1}(0)$.

Using the known Wigner function of the Fock state
$|n\rangle\langle n|$ \cite{Groen,DM85}
\[
W_n(p,q)=2(-1)^n \exp\left(-p^2-q^2\right)L_n\left(2p^2+2q^2\right)
\]
($L_n(x)$ being the Laguerre polynomial) and
the generating function of the Laguerre polynomials \cite{Bate}
\[
\sum_{n=0}^{\infty} \frac{z^n}{n!}L_n(x)=e^z J_0\left(2\sqrt{xz}\right),
\]
it is not difficult to write an
explicit expression for the Wigner function of the mixed state
$\widehat\rho=\sum p_n|n\rangle\langle n|$ with the coefficients given by
Eq. (\ref{p-evod}):
\begin{eqnarray}
W(p,q;\beta,r)&=&\frac{\exp\left(-p^2-q^2\right)}{\sinh(2r)}\left\{
\left[1-(1-2\beta)e^{-2r}\right] I_0\left(\sqrt{8r\left(p^2+q^2\right)}\right)
\right.\nonumber\\& +& \left.
\left[(1-2\beta)e^{2r}-1\right] J_0\left(\sqrt{8r\left(p^2+q^2\right)}
\right)\right\}.
\label{WIJa}
\end{eqnarray}
Here $J_0(z)$ is the Bessel function and $I_0(z)$ is the modified
Bessel function.
The Wigner function (\ref{WIJa}) has zero mean values of the quadratures
$q$ and $p$, and it is very different from the Wigner functions of the
pure even/odd coherent states
\begin{eqnarray}
W_\pm(p,q;\bar{p},\bar{q})&=&2N_\pm^2\Big\{
\exp\left[-\left(q-\bar{q}\right)^2 -\left(p-\bar{p}\right)^2 \right]
+\exp\left[-\left(q+\bar{q}\right)^2 -\left(p+\bar{p}\right)^2 \right]
 \nonumber\\ &\pm&
2 \exp\left(-q^2 -p^2\right)\cos\left[2\left(q\bar{p} -p\bar{q}\right)
\right]\Big\},
\label{pureWig}
\end{eqnarray}
where $\bar{p},\bar{q}$ are the mean values of the quadratures in the
initial coherent state $|\alpha\rangle$ with
$\alpha=\left(\bar{q}+i\bar{p}\right)/\sqrt2$.
However, assuming $\bar{q}=\sqrt{2r}\cos\varphi$,
$\bar{p}=\sqrt{2r}\sin\varphi$ and averaging $W_\pm(p,q;\bar{p},\bar{q})$
over the angle $\varphi$ according to the formula
\[
\widetilde{W}_\pm(p,q;r)\equiv \int_0^{2\pi}
\frac{d\varphi}{2\pi} W_\pm(p,q;\sqrt{2r}\cos\varphi,\sqrt{2r}\sin\varphi)
\]
we arrive exactly at Eq. (\ref{WIJa}) with $\beta=0$ for the even states and
$\beta=1$ for the odd states. For this reason we call the state described
by the Wigner function (\ref{WIJa}) as the {\it phase-averaged even/odd
state\/} (PAEOS). The phase-averaged {\it coherent states\/} were considered
in \cite{QE} in connection with the problem of a classical limit for the
quantum oscillator.
The PAEOS are quantum mixtures, since
the {\it purity coefficient\/}
\[
\mu\equiv \mbox{Tr}(\rho^2)
=\frac12\left\{\left(\frac{1-\beta}{\cosh r}\right)^2
\left[I_0(2r)+J_0(2r)\right] +\left(\frac{\beta}{\sinh r}\right)^2
\left[I_0(2r)-J_0(2r)\right]\right\}
\]
is less than $1$ for $r>0$. It is a monotonous function of $r$, whose
asymptotics are
\[
\mu\approx (1-\beta)^2(1-r^2)+\beta^2(1-r^2/3), \quad r\ll 1, \qquad
\mu\approx \left[(1-\beta)^2+\beta^2\right]/\sqrt{\pi r}, \quad
r\gg 1.
\]
Nonetheless, PAEOS are nonclassical states, since the Wigner function
$W(q,p)$ (\ref{WIJa}) assumes negative values, as shown in Figs 1 and 2,
where we plot $W(q,p)$ as a function of
$x=\sqrt{q^2+p^2}$. If $r>1$, then the plots corresponding to parameters
$1-\beta$ and $\beta$ have a mirror symmetry with respect to the
$x$-axis for $x<\sqrt{r/2}$, since in this region the
contribution of the oscillating
function $e^{r}J_0\left(x\sqrt{8r}\right)$ is dominating (note that
$W(0,0;r,\beta)=2(1-2\beta)$ does not depend on $r$). However, the
dependence on $\beta$ disappears for $x>\sqrt{r/2}>1$, where the
Wigner functions are close to zero in a wide interval (which increases with
an increase of $r$), then exhibit wide and not very high maxima
(at $x\approx\sqrt{2r}$), and
finally tend to zero exponentially for $x\gg\sqrt{2r}$.
In the special case $\beta=1/2$ the
Wigner function (\ref{WIJa}) is positive and does not
oscillate even for large values of the parameter $r$, as shown in Fig. 3.
Note that the purity coefficient $\mu$ also attains its minimum (for a
fixed value of $r$) when $\beta$ is close to $1/2$.

The type of the photon statistics (sub- or super-Poissonian) is determined by
Mandel's parameter ${\cal Q}\equiv {\cal N}_2 /{\cal N}_1 - {\cal N}_1$.
In the case of PAEOS this parameter equals
\[
{\cal Q}= \frac{r}{B}\left(1-B^2\right), \quad
B=(1-\beta)\tanh(r)+\beta\coth(r),
\]
so the photon statistics becomes sub-Poissonian for
$\beta>\frac12\left(1-e^{-2r}\right)$, i.e. $1-2\beta<e^{-2r}$.
In particular, in the case $\beta=1/2$ (Fig.~3) we still have the
sub-Poisson statistics.

Till now we assumed that we had no one-photon processes at all.
Now let us allow a {\it small\/} (but nonzero) coefficient $\nu$ (weak
one-photon processes).
Then we can simplify Eq. (\ref{sol-kum}) with the aid of the relation
$\lim_{\nu\to 0} \Phi(a\nu;c\nu;x)=
1+(a/c)\left(e^x -1\right)$.
In this limit, $R\to 2r$, $h\to r$, $g\to \frac12(1+s) +(s+\sigma)/(2r)$,
we arrive again at Eq. (\ref{F-evod}). The essential difference is that now
the coefficient $\beta$ is determined not by the initial conditions,
but by the relative strengths of the emission and absorption processes:
\begin{equation}
\beta=\sinh(r)\frac{\sinh(r) +(S/r)\cosh(r)}{\cosh(2r) +(S/r)\sinh(2r)},
\quad S\equiv \frac{s+\sigma}{s+1},
\label{beta-r}
\end{equation}
and it is always less than $1/2$, since
\[
1-2\beta=\left[ \cosh(2r) +(S/r)\sinh(2r)\right]^{-1} >0.
\]
It is remarkable that parameter $\nu$ does not enter the formulas
describing the stationary distribution in the limit $\nu\ll 1$.
It influences only
the relaxation time $t_{rel}\sim \nu^{-1}$, but not the form of the
stationary state.
If $r\to 0$ (no two-photon emission, two-photon absorption only),
then $\beta\to S/(1+2S)$. If we have
no one-photon emission ($S\to 0$), then $\beta \to \tanh^2(r)/
\left(1+ \tanh^2(r)\right)$. The maximal value $\beta=\frac12$ is achieved
when $S\to\infty$ or $r\to\infty$.

Mandel's parameter reads now
\begin{equation}
{\cal Q}=\frac{r\left[1-(S/r)^2\right]\left[1-\tanh^2(2r)\right]}
{[1+(S/r)\tanh(2r)][(S/r)+\tanh(2r)]}.
\label{Q-r}
\end{equation}
The photon statistics is sub-Poissonian if $r<S$, and super-Poissonian
if $r>S$. For a fixed $r$, function ${\cal Q}(S)$ monotonously
decreases from ${\cal Q}(0)=r\coth(2r)\left[1-\tanh^2(2r)\right]$ to
${\cal Q}(\infty)=-r\coth(2r)\left[1-\tanh^2(2r)\right]$. Consequently,
$-\frac12<{\cal Q}(r,S)<\frac12$.

\section{Conclusion}

Let us formulate the main results of the paper. We have found an exactly
solvable 6-parameter family of stationary master equations for the
diagonal elements of the 1-mode field in a cavity in the presence of
competing one- and two-photon emission and absorption processes, and we
gave explicit solutions for its 4-parameter subfamily describing the
completely saturated two-photon emission regime. We have shown that in the
limit case of weak one-photon processes, the field mode goes to the
nonclassical stationary state which can be considered as a mixed analog
of even and odd pure coherent states. Although we considered an idealized
case of a {\it completely saturated\/} two-photon emission, the results
obtained could help to understand the qualitative features of real
(partially saturated) processes.

\medskip

\noindent {\bf Acknowledgements}
This research was supported by FAPESP, project 1995/3843-9.
SSM thanks CNPq and FINEP, Brasil, for partial financial support.

\vspace{2cm}

{\bf Figure 1.} Wigner function $W(x)$, $x\equiv\sqrt{q^2+p^2}$,
of the phase-averaged
{\it even\/} state ($\beta=0$) for $r=10$.

\vspace{2cm}
{\bf Figure 2.} Wigner function $W(x)$, $x\equiv\sqrt{q^2+p^2}$,
of the phase-averaged
{\it odd\/} state ($\beta=1$) for $r=10$.

\vspace{2cm}
{\bf Figure 3.} Wigner function $W(x)$, $x\equiv\sqrt{q^2+p^2}$,
of the phase-averaged
``maximally mixed state'' ($\beta=0.5$) for $r=10$.

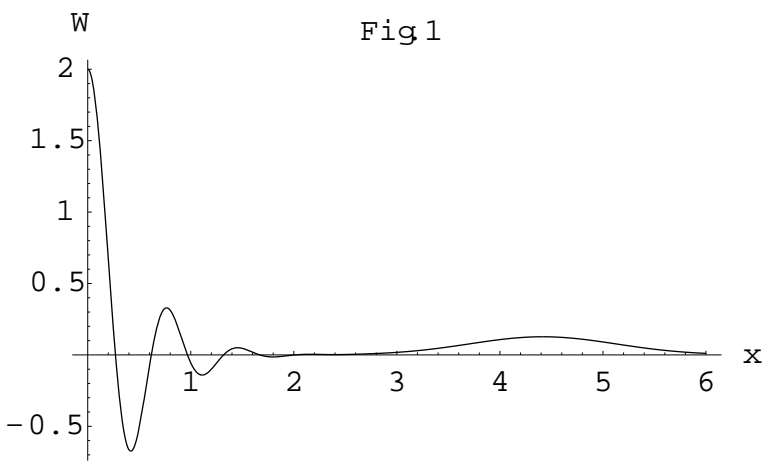
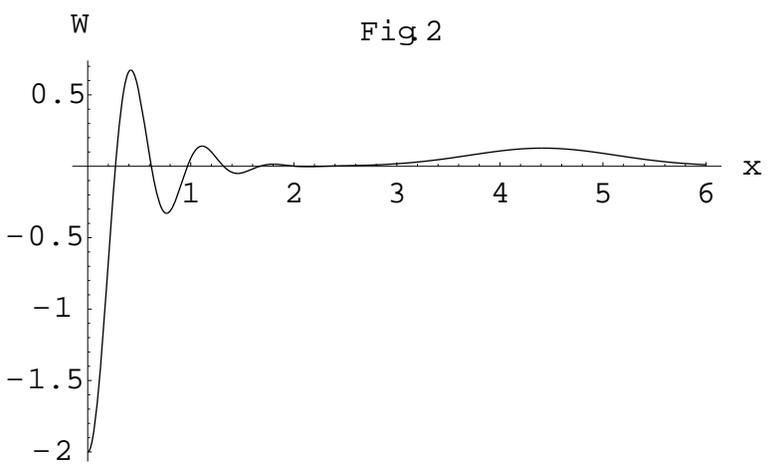
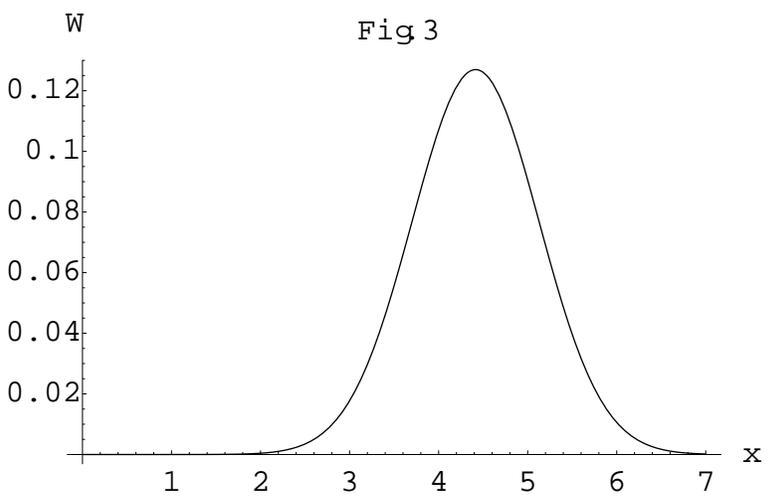
\end{document}